\documentclass[%
superscriptaddress,showkeys,
showpacs,preprintnumbers,
 amsmath,amssymb,
 aps,
pra,
twocolumn,
]{revtex4-1}
\usepackage{graphicx}% Include figure files
\usepackage{dcolumn}% Align table columns on decimal point
\usepackage{bm}% bold math
\usepackage{hyperref}% add hypertext capabilities
\usepackage{natbib}
\usepackage{color}
\usepackage[ansinew]{inputenc}

\begin{document}

\title{Effects of temperature and ground-state coherence decay \\ on enhancement and amplification in a Delta atomic system}
\author{Manukumara Manjappa}
\affiliation{%
	Raman Research Institute Bangalore, 560080, India
	}
\author{Satya Sainadh Undurti}
\affiliation{%
	Raman Research Institute Bangalore, 560080, India
	}
\author{Asha Karigowda}
\affiliation{%
	Raman Research Institute Bangalore, 560080, India
	}
\affiliation{%
	Department of Physics, Kuvempu University,  Shivamogga,  577451, India
	}
\author{Andal Narayanan}
\email{andal@rri.res.in}
\affiliation{%
	Raman Research Institute Bangalore, 560080, India
	}
\author{Barry C. Sanders}
\email{sandersb@ucalgary.ca}
\affiliation{%
    Institute for Quantum Science and Technology, University of Calgary, Alberta, Canada T2N 1N4
    }
\affiliation{%
    Program in Quantum Information Science,
    Canadian Institute for Advanced Research,
    Toronto, Ontario M5G 1Z8, Canada
    }
\affiliation{%
	Hefei National Laboratory for Physical Sciences at Microscale and Department of Modern Physics,
	University of Science and Technology of China, Hefei, Anhui 230026, China
    }
\affiliation{%
	Shanghai Branch,
	CAS Center for Excellence and Synergetic Innovation Center
		in Quantum Information and Quantum Physics,
	University of Science and Technology of China, Shanghai 201315, China
	}

\date{\today}


\begin{abstract}
We study phase-sensitive amplification of electromagnetically induced transparency in a warm $^{85}$Rb vapor wherein a 
microwave driving field couples the two lower energy states of a $\Lambda$ energy-level system thereby transforming into a 
$\Delta$ system.  Our theoretical description includes effects of ground-state coherence decay and temperature effects.  In particular,
we demonstrate that driving-field enhanced electromagnetically induced transparency is 
robust against significant loss of coherence between ground states.
We also show that
for specific field intensities, 
a threshold rate of ground-state coherence decay exists at every temperature.
This threshold separates the probe-transmittance behavior into two regimes:
probe amplification vs.\  probe attenuation.
Thus,
electromagnetically induced transparency plus amplification is possible at any temperature in a~$\Delta$ system.
\end{abstract}

%\keywords{latex-community, revtex4, aps, papers}
\maketitle

\section{Introduction \label{intro}}
\begin{figure}
\includegraphics[scale=.60]{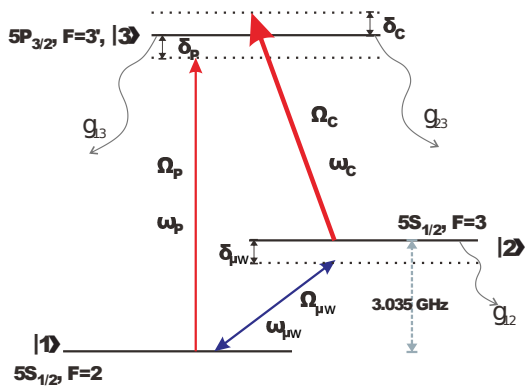}
\caption{(Color online) Level scheme showing~$\Delta$ atomic level configuration in $^{85}$Rb formed by two 
optical fields and a microwave field.
The coupling field at frequency~$\omega_\text{c}$,
with strength ~$\Omega_\text{c}$ (thick red),
connects the $5S_{1/2}, F=3~(|2\rangle) \leftrightarrow 5P_{3/2}, F=3'~(|3\rangle)$ transition.
The probe field at frequency~$\omega_{\text{p}}$ (thin red),
with strength~$\Omega_{\text{p}}$, connects the 
$5S_{1/2}, F=2~(|1\rangle) \leftrightarrow 5P_{3/2}, F=3'~(|3\rangle)$ transition.
The microwave drive field at frequency~$\omega_{\mu \text{w}}$, with strength
$\Omega_{\mu \text{w}}$ connects levels $5S_{1/2}$
$F=2~(|1\rangle) \leftrightarrow 5S_{1/2}$, $F=3~(|2\rangle)$.
The detunings of the coupling and probe fields
from $|3\rangle$ are denoted by $\delta_{\text{c}}$ and~$\delta_{\text{p}}$ respectively. 
The decay rate~$\gamma_{ij}$
from level $|j\rangle$ to level $|i\rangle$
assumes values $\gamma_{12} = 0.001$~MHz and~$\gamma_{23} = 5$~MHz.
\label{fig:level_diagram}}
\end{figure}

Electromagnetically induced transparency (EIT)~\citep{harris-prl-62-1033} has become 
foundational for 
creating, storing and transfering quantum features between interacting systems.  EIT has its origins in atom-light interaction
phenomena wherein an atom with a three-level $\Lambda$ configuration of atomic levels
interacts with two coherent electromagnetic fields ($|1\rangle \leftrightarrow |3\rangle$ and 
$|2\rangle \leftrightarrow 3\rangle$ of Fig.~\ref{fig:level_diagram}). 
Under a two-photon
resonance condition for these two transitions, absorption for both these fields is eliminated due to
establishment of quantum coherence between levels~$|1\rangle$ and~$|2\rangle$.  Several other non-intuitive physical phenomena arise due to this 
quantum coherence.  These include lasing without inversion (LWI)~\citep{prl-62-2813},
the realization of slow and stopped light~\citep{nature-409-490},
ultra-low light level optical switches~\citep{pra-68-041801R} and single-photon quantum 
non-linearities~\citep{science-333-1266}.\\
\indent For a long time~\citep{opt-acta-33-1129,opt-comm-85-209}, it was realized that connecting the unconnected two levels
of a traditional EIT system by a drive field makes the absorption and dispersion properties dependent on the relative
phase between all three fields. For a $\Lambda$ EIT system,this would result in connecting the lower 
two levels~$|1\rangle$ and~$|2\rangle$,
resulting in a~$\Delta$ system (Fig.~\ref{fig:level_diagram}). From an analysis of dark states in a $\Lambda$ system controlled by
a microwave field, the sensitivity of the dark state to the relative phase between the interacting fields was brought out~\citep{jphysb-42-235505}.
Spatially seperated interaction with the Raman optical fields and the microwave field translated the optical dark
state to either one of the microwave-dressed spin states and vice-versa~\cite{prl-65-1865}.
A study of slow and fast light propagation in a 
$\Delta$ system~\citep{pra-64-053809} explicitly brought out the control on dispersive properties of such systems by
the drive field. Experimental demonstration of both 
EIT~\citep{pra-72-063813,pra-80-023820} and associated non-linear effects~\citep{epl-95-34005}
in~$\Delta$ systems of~$^{85}$Rb vapor
has opened the possibility of demonstrating phase-sensitive coherence-related effects
in these systems.\\
\indent Phase-dependent amplification for microwave fields in a
fluxonium
superconducting circuit with an artificial atom featuring a~$\Delta$ configuration of energy levels
reveals simultaneous existence of
LWI and EIT phenomena resulting in probe amplification~\citep{prl-105-073601}. 
As
 the propagation phases of electromagnetic (EM) fields governed by Maxwell equations depend on the polarization of the medium,
the EM waves participating in a closed-loop interaction scheme, 
thus have their absorptive and dispersive properties determined not
only by Kramers-Kronig relations but also by the refraction experienced in the medium~\citep{pra-59-2302}.
Hence,
for approriate phases, when the medium as a whole exhibits transparency for all three interacting fields in a~$\Delta$ system,
the total EM energy can oscillate between any of these fields.
This oscillation can give rise to large lossless amplification  in any one or
two of the three fields~\citep{pra-60-4996}. An analysis of phase dependent amplification of a probe pulse
controlled by the drive-field~\cite{opt-lett-34-1834} used amplification to compensate for losses experienced inside
the medium. Fourier decomposition of pulse propagation in a $\Delta$ system~\citep{josab-30-1517}
showed the simulatenous existence of absorption and gain channels due to multi-photon processes.\\
\indent Despite such strong reasons to observe amplification in~$\Delta$ systems, 
lossless amplification of  probe field has not been experimentally verified so far,
even though enhancement of EIT has been seen~\citep{pra-80-023820}.  Factors
that affect population and coherence between the ground states of a~$\Delta$ system play a major role in determining such experimental
outcomes.  For room-temperature experiments, the presence of an associated thermal bath determines steady-state population between the ground 
levels.
In addition, various factors affect the coherence between the ground states of a~$\Delta$ system.\\
\indent One such factor is 
the finite bandwidth of the optical and microwave drive fields.  For finite bandwidth electromagnetic fields, the phase diffuses
over time across their bandwidths.  
For a $\Lambda$ system, with critically cross-correlated probe
and coupling fields,
such phase diffusion does not destroy ground-state atomic coherence 
and preserves the dark state~\citep{jphysb-15-3997-1982}.
In a~$\Delta$ system,
the issue of phase diffusion on coherence between ground sates has not been addressed so far.
However, Agarwal's treatment
of~$\delta$-correlated phase fluctuating fields interacting with two-level atoms~\citep{pra-18-1490-1978}
suggests that any formal treatment
of phase fluctuations in a~$\Delta$ configuration will lead to ground-state coherence decay.
In addition to phase fluctuations,
collisions between Rb atoms and between the Rb atoms and the walls of the cell
also contribute to decay of ground-state coherence.\\
\indent Here we undertake a comprehensive analysis of 
the consequences of thermal bath and the effect of decay rate of ground-state coherence on a~$\Delta$ atomic system.
For the purpose of this study,
we have used the natural decay parameters pertaining to a realistic~$\Delta$ system in a warm $^{85}$Rb vapor.  In addition, we have
treated the rate of ground-state coherence decay as a phenomenological constant.
This constant represents the various physical effects
that give rise to a decay of ground-state coherence in a~$\Delta$ system. Collisional factors that give rise to ground-state coherence decay
can be mitigated by using buffer gas additions to Rubidum vapors and by using paraffin coated vapor cells. Therefore, our present study is 
most relevant to ground-state coherence decay given rise by the ever present finite band-width of electromagnetic fields used in the experiment. 
In this context, our study helps to quantify the robustness of phase-sensitive induced-transparency effects, with phase diffusing fields.\\
\indent For given intensities of coupling, probe and microwave fields,
we establish the existence of a threshold rate of ground-state coherence decay at every temperature.  This threshold separates the behavior into regimes.
Below this threshold,
probe enhancement and amplification are possible 
for a wide range of coherence decay values.
Importantly, our analysis establishes that enhancement of the probe field in the presence of a drive field
is a precursor to probe amplification.
Furthermore, transparency and amplification are possible even for warm~$\Delta$ systems, provided
the rate of ground-state coherence decay is below a certain limit.\\
\indent The outline of our paper is as follows.  Section II details our theory of~$\Delta$ systems which includes thermal bath and 
ground-state coherence decay effects.  Section
III  presents results of our theoretical model. Section IV discusses the results and gives predictions for future 
experiments. We summarize our conclusions in section V.
\section{Theory \label{theory}}
The theory of our atomic~$\Delta$ system differs from other theories of closed-loop systems~\citep{pra-63-043818-2001}
in two respects. In contrast to previous analyses,
we have included the consequences of decoherence between the lower two levels 
$|1\rangle$ and~$|2\rangle$ by making the ground-state coherence decay rate as a variable in our calculations. In addition, we have 
included thermal bath effects for finite temperature~$\Delta$ systems.\\
\indent We consider the atomic levels of our~$\Delta$ system as shown in Fig.~\ref{fig:level_diagram}, which also presents interactions with the probe~(p), coupling~(c) 
and microwave~($\mu$w) fields. The Rabi frequency $\Omega$ for dipole interaction of a pair of levels~$|i\rangle$ and~$|j\rangle$, with an applied field
$\bm{E}$, is given by $\Omega = \bm{d}_{ij} \cdot \bm{E}$, with~$\hbar$ taken to be equal to 1, and~$\bm{d}_{ij}$ $\equiv$ $\langle i|\bm{d}|j\rangle$
with~$\bm{d}$ the dipole moment vector.\\
\indent For our~$\Delta$ system, we take the optical probe and coupling fields to be
propagating through the cell along the~$z$ axis.
They are represented by
\begin{equation}
	\bm{E}_{\perp \text{p}} (\bm{r}_{\perp}) \cos(\omega_\text{p}t-k_{\text{p}}z+\phi_\text{p})
\end{equation}
and
\begin{equation}
	\bm{E}_{\perp \text{c}} (\bm{r}_{\perp}) \cos(\omega_\text{c}t-k_{\text{c}}z+\phi_\text{c}),
\end{equation}
with angular frequencies~$\omega_\text{p,c}$, wave numbers~$k_\text{p,c}$ and initial phases~$\phi_\text{p,c}$ respectively. We have taken the microwave field 
to be a standing wave inside a microwave cavity;
therefore there is no propagation phase associated with it.
The microwave field is thus
represented by 
\begin{equation}
	\bm{E}_{\mu\text{w}}(\bm{r})\cos(\omega_{\mu\text{w}}t+\phi_{\mu\text{w}}).
\end{equation}

\indent We start with 
the Hamiltonian of the~$\Delta$ system in the interaction and rotating wave picture, taking the reference energy
level as the energy of level $|3\rangle$:
\begin{align}
	\hat{H}(\bm{r},\bm{v}) = \delta'_\text{p} (\bm{v}) \vert 1\rangle \langle 1\vert + 
\delta'_\text{c} (\bm{v}) \vert 2\rangle \langle 2\vert +
\Omega_{\mu\text{w}}(\bm{r})\vert 1\rangle \langle 2\vert \nonumber \\
 +\Omega_\text{p}(\bm{r}_\perp)\vert 1\rangle \langle 3\vert + \Omega_\text{c}(\bm{r}_\perp)\vert 2\rangle
 \langle 3\vert + \text{h.c.}
\label{eq:a1}
\end{align}
for h.c.\ denoting Hermitian conjugate.
Here
\begin{equation}
\label{eq:delta'}
	\delta'_\text{p} (\bm{v}) \equiv \delta_\text{p}- \bm{k}_\text{p} \cdot \bm{v},\; 
	\delta'_\text{c} (\bm{v}) \equiv \delta_\text{c}- \bm{k}_\text{c} \cdot \bm{v},
\end{equation}
are Doppler-shifted detunings of the coupling and probe fields (see Fig.~\ref{fig:level_diagram}) 
seen by an atom moving with velocity $\bm{v}$. The terms~$\Omega_j$'s are complex Rabi frequencies of the fields are 
\begin{align}
	\Omega_\text{c} (\bm{r}_\perp)
		:=&\bm{d}_{23} \cdot \bm{E}_{\perp \text{c}} (\bm{r}_\perp) \text{e}^{\text{i}\phi_\text{c}}, \nonumber\\
	\Omega_\text{p} (\bm{r}_\perp)
		:=&\bm{d}_{13} \cdot \bm{E}_{\perp \text{p}} (\bm{r}_\perp) \text{e}^{\text{i} \phi_\text{p}},
\nonumber\\
	\Omega_{\mu\text{w}}(\bm{r})
		:=&\bm{d}_{12} \cdot \bm{E}_{\mu\text{w}}(\bm{r}) 
\text{e}^{\text{i} (\omega_\text{p}-\omega_\text{c}-\omega_{\mu\text{w}})t+\text{i}(k_\text{p}-k_\text{c})z 
	+\text{i}\phi_{\mu\text{w}}}.
\end{align}
Using the dipole approximation,
the Rabi frequencies of the probe,
coupling,
and microwave fields are taken to be spatially uniform, yielding the constants
$\Omega_\text{c} (\bm{r}_\perp)\equiv\Omega_{\text{c}}$,
$\Omega_\text{p} (\bm{r}_\perp)\equiv\Omega_{\text{p}}$
and $\Omega_{\mu\text{w}}(\bm{r})\equiv\Omega_{\mu\text{w}}$.

Unlike a 
$\Lambda$ system, the propagation and temporal
phases of the EM fields in closed-loop systems do not vanish in the interaction picture.
Choosing $\delta_\text{c}$ = 0, and maintaining
$\delta_\text{p} = \delta_{\mu\text{w}}$ in
all our calculations,
makes the temporal phase factor
\begin{equation}
	\omega_\text{p}-\omega_\text{c}-\omega_{\mu\text{w}} = 0,
\end{equation}
thus ensuring time independent
Rabi frequencies.
However, the propagation phases give rise to an effective position-dependent microwave Rabi frequency, which is given by
$\Omega_{\mu \text{w}}(z) = \Omega_{\mu\text{w}}\text{e}^{\text{i}\phi_{\mu\text{w}}}  \text{e}^{\text{i}(k_\text{p}-k_\text{c})z}$.

\indent The dynamical evolution of density matrix elements, in the interaction picture is given 
by the master equation
\begin{align}
	\dot{\rho}(\bm{v},z)
		=- i[\hat{H}(\bm{v},z),\rho(\bm{v},z)]+\sum_{k=1}^5\mathcal{L}(\hat{c}_k)\rho(\bm{v},z)   \label{eq:a2}
\end{align}
with~$\mathcal{L}(\hat{c}_k)$ being the Lindblad superoperator
\begin{equation}
	\mathcal{L}(\hat{c})\rho:=\hat{c}\rho\hat{c}^\dagger-\frac{1}{2} 
		\{\rho,\hat{c}^\dagger\hat{c}\}
\end{equation}
acting on operators
\begin{align} 
	\hat{c}_1=&\sqrt{(\bar{n}+1)\gamma_{12}}\vert 1\rangle \langle 2\vert, \nonumber \\
	\hat{c}_2=&\sqrt{\bar{n}\gamma_{12}}  \vert 2\rangle \langle 1\vert,\;
	\hat{c}_3=\sqrt{\gamma_{13}}  \vert 1\rangle \langle 3\vert, \nonumber \\
	\hat{c}_4=&\sqrt{\gamma_{23}} \vert 2\rangle \langle 3\vert,\;
	\hat{c}_5=\sqrt{\gamma_{\text{c}}}  (\vert 1\rangle \langle 1\vert-\vert 2\rangle \langle 2\vert) \label{eq:a3}
\end{align} 
with~$\gamma_{12}$ and $\gamma_{23}=\gamma_{13}$ representing the natural linewidth of levels~$\vert 2 \rangle$ and 
$\vert 3 \rangle $ respectively.

The symbol~$\bar{n}$ 
is the average number of thermal photons in the bath at temperature~$T$,
and~$\gamma_{\text{c}}$ is the rate of ground-state coherence decay. As our optical fields are co-propagating along the~$z$ direction, 
we henceforth denote the~$z$ component
of the velocity vector by $v$. We solve Eqs.~(\ref{eq:a1}-\ref{eq:a3})
for steady-state values of~$\rho(v)$ which is then 
averaged over the Maxwell-Boltzmann velocity profile at some temperature~$T$  
\begin{equation}
	\bar{\rho}
		= \dfrac{\int_{-\infty}^\infty \rho(v)\text{e}^{-\left(\frac{v}{v_\text{mp}}\right)^2}\text{d}v}{\int_{-\infty}^\infty \text{e}^{-\left(\frac{v}{v_\text{mp}}\right)^2}\text{d}v} 
\label{eq:vel-ave}
\end{equation} 
with~$v_{\text{mp}}=\sqrt{\frac{\pi}{4}}$ $\bar{v}$ the most probable speed of atoms at temperature~$T$ 
and~$\bar{v}$ the average speed of the atoms.\\
\indent As is well known in a~$\Delta$ system, the steady-state $\bar{\rho}$ matrix values depend on the relative
phase between all fields~\citep{opt-acta-33-1129}.
As the coupling and probe fields differ in wavelength, they have differing
phase values during propagation. We take both these optical fields to be derived from the same source thereby making their 
initial phases identical. Thus, the relative phase between all three fields is
\begin{equation}
	\phi(z) = z(k_\text{p}-k_\text{c}) + \phi_{\mu\text{w}}. \label{eq:phi-def}
\end{equation}
Therefore, we can vary the value of~$\phi$ by controlling~$z$.
As experiments with~$\Delta$ systems typically employ a Rb vapor cell of a finite length~$L$,
$z$-dependent phase variations of density-matrix elements need to be calculated, taking into account the 
phase changes experienced over the entire length~$L$. In addition, we have assumed that the probe and coupling fields
are right and left circularly polarized, reflecting experimental demonstration of probe transmission sensitivity
to polarizations~\citep{pra-80-023820}.

In order to simulate the changes in the probe  field as it passes through a cell of length~$L$,
we treat the Rb cell as a sequence of small cells along 
the propagation direction. The propagation equation for the probe field is then calculated using the slowly-varying envelope approximation in each cell,
which is given by~\citep{pra-80-023820}
\begin{equation}
	\frac{\partial \Omega_\text{p}}{\partial z}=-i \eta \bar{\rho}_{_{31}}.
\label{eq:prop}
\end{equation}
Here $\bar{\rho}_{31}$ represents the phase-dependent steady-state density matrix element corresponding to probe absorption,
obtained using Eqs.~(\ref{eq:a1}-\ref{eq:a3}),
and~$\eta$ is the coupling constant taken to be close to 1.\\
\indent In the following section, we present results for probe transmission from our~$\Delta$ system using our theoretical
model. The absorption experienced by the probe field during interaction
with an  $^{85}$Rb atom at a position~$z$, are calculated using the imaginary parts of the density matrix element
$\bar{\rho}_{31}$. Using realistic parameters, we present results of change in probe transmission for a system of  $^{85}$Rb atoms contained in a vapor 
cell of length~$L$, for various values of~$T$ and  for various rates of ground-state coherence decay ($\gamma_{\text{c}}$) using
Eq.~(\ref{eq:prop}).

It is well known that ground-state coherence decay rate in a $\Lambda$ system affects the contrast of EIT transmission resonance. In subsequent
sections, we explore the transmission loss in the probe beam of our~$\Delta$ system as a function of~$T$ and~$\gamma_{\text{c}}$, both of
which affect the coherence between the ground states.
\section{Results}
\begin{figure}
\includegraphics[width=0.8\columnwidth]{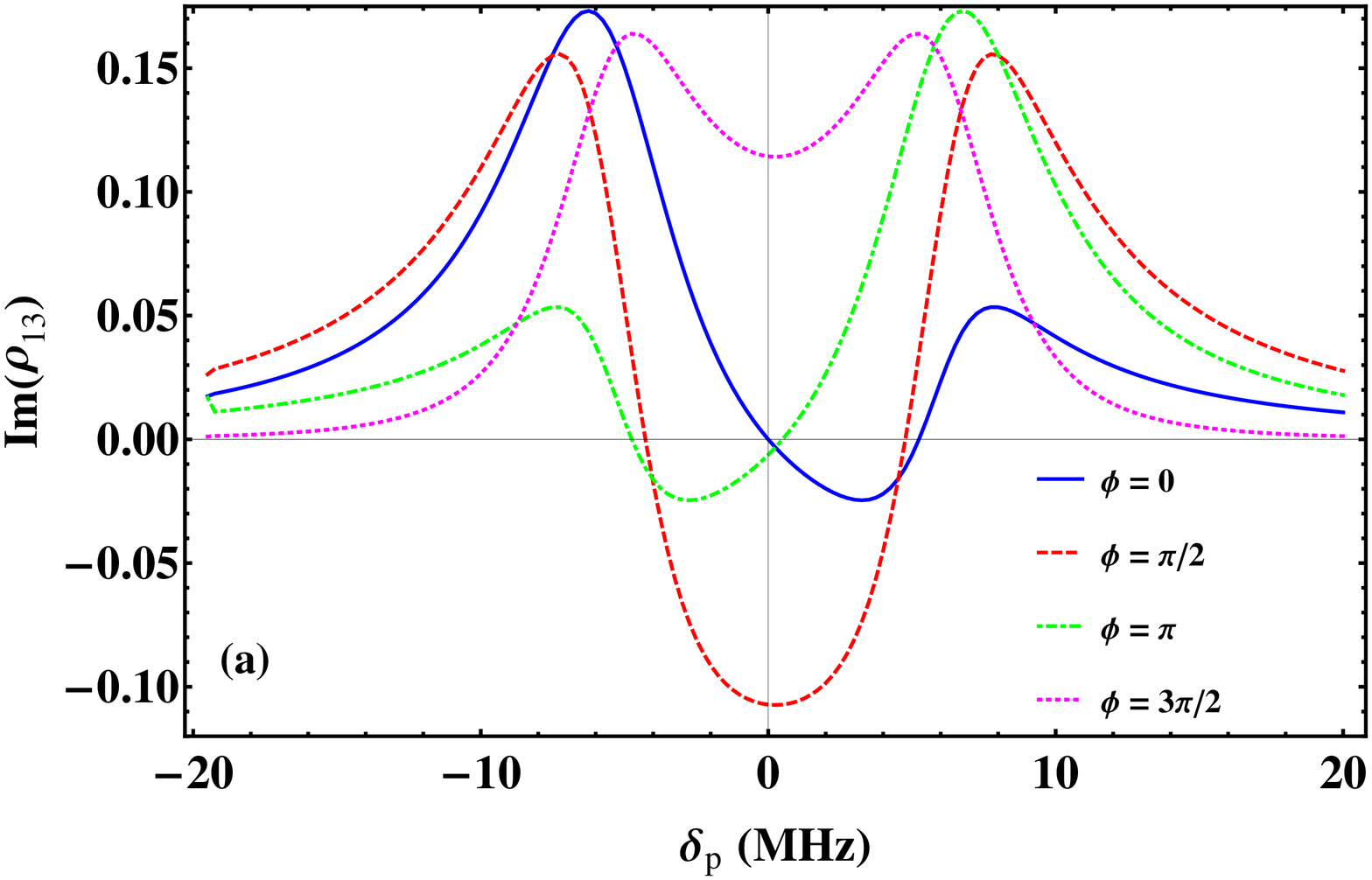}\\
\includegraphics[width=0.8\columnwidth]{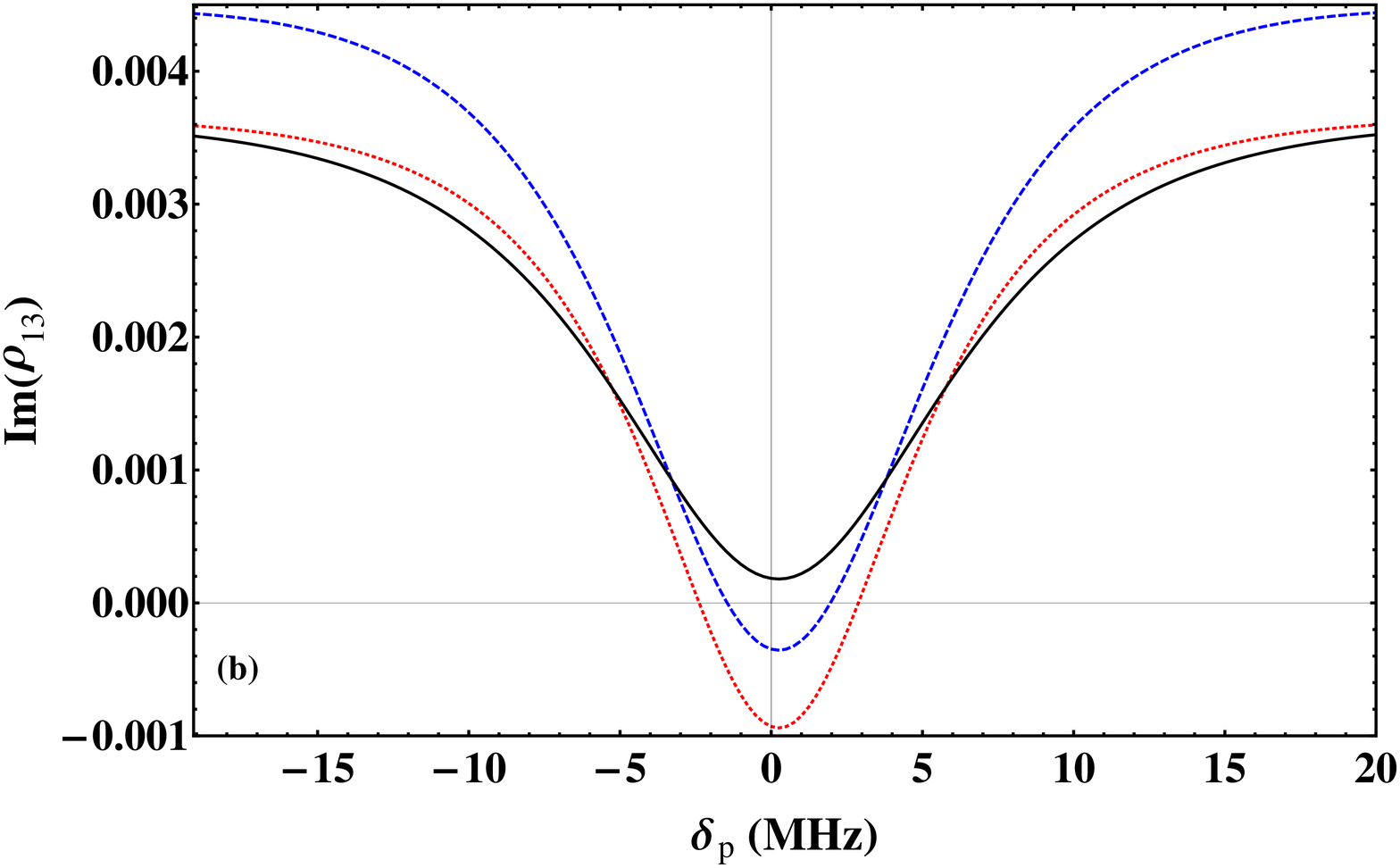}
\caption{
	(Color online)
	\color{black}{ Probe absorption~Im$(\rho_{13})$ vs.\ probe detuning~$\delta_\text{p}$
	for a $\Delta$ system in $^{85}$Rb gas with
	$\Omega_\text{c}=6.4$~MHz,
	$\Omega_\text{p}=1.0$~MHz,
	and $\Omega_{\mu\text{w}}=0.8$~MHz.
	(a)~Probe absorption with no ground-state coherence decay ($\gamma_{\text{c}} = 0$~MHz)
        at $T$ = 0~K is given. The various values of~$\phi$ are indicated in the legend.
	(b)~Probe absorption with finite ground-state coherence decay values and 
        at finite temperatures given for a fixed value of $\phi = \pi/2$. The 
        various cases are:
	$T=333$~K and $\gamma_{\text{c}}=1.0$~MHz
	(solid black line),
	$T=333$~K and $\gamma_{\text{c}}=0.1$~MHz
	(dot-dashed red line), and
	$T=233$~K and $\gamma_{\text{c}}=1.0$~MHz.
	(dashed blue line)
        }
	}
\label{fig:absorption0K}
\end{figure}

In Fig.~\ref{fig:absorption0K}(a)
we show probe absorption as a function of~$\delta_\text{p}$
for an atomic~$\Delta$ system at position~$z$ held at a temperature~$T=0$~K with 
no coherence decay between the ground states ($\gamma_{\text{c}} = 0$~MHz). 
Changing~$z$ results in modified values of relative phase $\phi$. We see from the figure that
for $\phi =  \pi/2$ there is significant amplification seen in the probe field around two-photon resonance ($\delta_\text{p} = \delta_\text{c}$)
Thus Fig.~\ref{fig:absorption0K}(a)
establishes that our $\Delta$-system theory produces phase-sensitive probe amplification at $T=0$~K in the limit
of zero rate of ground-state coherence decay.
We also present in Fig.~\ref{fig:absorption0K}(a)
the asymmetric and absorptive probe profiles at other $\phi$ values,
which are qualitatively similar to those seen in theoretical calculations of a~$\Delta$ system in superconducting circuits~\citep{prl-105-073601}.\\
\indent The effect increasing the ground-state coherence decay rate and temperature on probe transmission is revealed by the plots in Fig.~\ref{fig:absorption0K}(b), which pertain to $\phi=\pi/2$.

This~$\Delta$ system is assumed to be in a cell of length~$L$ with physical
parameters that give rise to a decay rate of ground-state coherence of about $1$~MHz. This is a realistic decay rate for experiments
conducted in narrow-diameter cells~\citep{ol-32-338-2007}.

With $\gamma_{\text{c}}=1.0$~MHz,
we see that
the probe experiences absorption at the two-photon resonance condition
for $\phi=\pi/2$.
For precisely this~$\phi$ value,
significant amplification of the probe transmission was obtained 
for parameters of Fig.~\ref{fig:absorption0K}(a).
Thus, increasing temperatures
and rates of ground-state coherence decay contributes towards a loss of probe amplification.

In Fig.~\ref{fig:absorption0K}(b),
the transmitted probe field does not endure absorption and is actually amplified at $T=333$~K
for $\gamma_\text{c}=0.1$~MHz.
Thus, we observe that non-zero values of~$\gamma_{\text{c}}$ can yield probe-field amplification
even for a warm~$\Delta$ system.

We compare absorption at $T=333$~K vs $T=233$~K
in Fig.~\ref{fig:absorption0K}(b), which shows that the transmitted probe in the cooler case is amplified.
As the drive-field intensities of all three graphs in~Fig.~\ref{fig:absorption0K}(b) are identical,
these plots illustrate that 
probe-field amplification can be obtained
for suitable values of the temperature~$T$
and the ground-state coherence decay rate~$\gamma_{\text{c}}$.

In Fig.~\ref{fig:probeintensitychange} we give a contour plot of change in transmitted probe intensity, 
as it emerges from a cell of length $L$ = 5 cm, for wide ranges of ground-state coherence decay rate values
and temperatures. The decay rate ranges from the kHz to the MHz domain, 
which incorporates regimes where collisional decay is the dominant decohering mechanism 
as well as regimes
where the finite linewidths of the electromagnetic fields contribute dominantly to the decay.
\section{Discussion}
\begin{figure}
\includegraphics[height=6.5 cm, width = 8.6 cm]{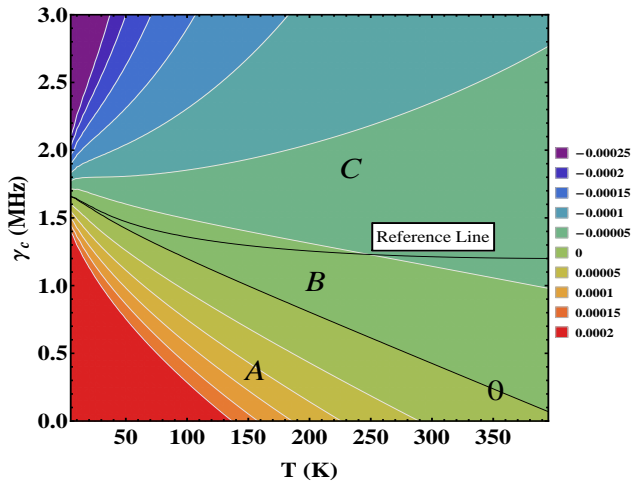}
\caption{
	(Color online)
	Contour plot showing the change from probe-field input intensity to transmitted intensity 
	for temperaturs $T=0$~K to $T=400$~K
	and for~$\gamma_{\text{c}}$ ranging from $0.001$~MHz to $3.000$~MHz.
	The Rabi frequencies are fixed at
	$\Omega_\text{c} = 6.4$~MHz, 
	$\Omega_\text{p} = 1.0$~MHz, and
	$\Omega_{\mu\text{w}} = 0.8$~MHz.
	`Reference Line' denotes a reference contour 
	along which the probe-field transmitted intensity is identical whether
	the microwave drive is on or off. 
	The symbols \emph{A}, \emph{B}, and \emph{C}
	refer to regions of amplification, enhancement and absorption respectively.
	}
\label{fig:probeintensitychange}
\end{figure}
\indent We see from the contour plot of Fig.~\ref{fig:probeintensitychange} that,
for given intensities of the coupling, probe and microwave fields, at
every temperature~$T$, there exists a 
threshold rate of ground-state decoherence $\gamma_{\text{th}}$. This threshold at every temperature lies along the contour denoted by zero in 
Fig.~\ref{fig:probeintensitychange}. For $\gamma_{\text{c}}$ values below the threshold value, we obtain
amplification in the optical probe field indicated by
positive-valued contours. For $\gamma_{\text{c}}$ values above this threshold the probe exhibits absorption signified by negative-valued 
contours.

Experimentally an increase of the transmitted optical probe intensity 
has been observed when the microwave field was
switched on compared to the value 
when the microwave field was switched off~\citep{pra-80-023820,epl-95-34005}.
The increased transmitted intensity was still lower than the input probe 
intensity, hence no amplification.

To understand this enhanced transmission of the probe field,
we show in the same 
contour plot of Fig.~\ref{fig:probeintensitychange} the special contour
labeled `Reference Line'.
Along this contour, the transmitted intensity of the probe field is the same,
with and without the microwave drive field on.
With the help of the `Reference Line' contour and the zero contour line, the contour plot can be divided into three regions.

Region~A of the contour plot is below the zero contour line, 
which is the region for which the probe field experiences amplification.
Region~B, which is sandwiched between the zero contour and the `Reference Line',
is where, 
despite absorption of the probe field,
increased transmittance occurs
compared to the absence of the microwave drive field.
Region~B is the region of enhancement.

Both of the~A and~B regions
are below the `Reference Line'.
Region~C is above the `Reference Line', gives the regime of probe absorption where the probe
experiences greater absorption than it did in the absence of the microwave drive field.\\
\indent Comparing $\gamma_{\text{c}}$ values along the `Reference Line' and the threshold $\gamma_{\text{th}}$ value for amplification along
the zero-valued contour, it is clear that $\gamma_{\text{th}} < \gamma_{\text{c}}$ at all values of temperature~$T$. This clearly indicates that
a loss in ground-state coherence is the main reason for absence of probe amplification in a~$\Delta$ system.

Significanlty, our result shows that enhancement 
is a precursor to amplification even for hot~$\Delta$ systems. This conclusion predicts that, in experiments with hot~$\Delta$ systems, if enhancement 
but not amplification is observed, then reducing factors which contribute to ground-state decoherence will
enable to obtain amplification. {\color{black}We have not considered intensity
dependent variations of susceptibility in this study, since the probe field amplification is small}.\\
\indent We give below an analytical 
derivation of~$\gamma_{\text{th}}$ at $T=0$~K to show its dependence on the intensities of coupling, probe and microwave fields.
At $T=0$~K, all the atoms are initially in the ground state $|1\rangle$ 
($\rho_{11} = 1$ and $\rho_{22} = 0$) and we assume $\rho_{23}\approx 0$ holds.

Using Eqs.~(\ref{eq:a1}-\ref{eq:a3}) 
we can solve for probe absorption in the medium,
using the steady-state $\rho_{31}$ expression given by
\begin{equation}
	\rho_{31}
		=\dfrac{i \Gamma^0_{12}\Omega_\text{p} -\Omega_\text{c} \Omega_{\mu\text{w}}
			\text{e}^{\text{i}\Delta kz}}{\Gamma^0_{12}\Gamma^0_{13}+|\Omega_\text{c}|^2}
\end{equation} 
with 
\begin{equation}
	\Gamma^0_{12}=\frac{1}{2}\gamma_{12}+2\gamma_{\text{c}}-\text{i}\Delta_{\mu\text{w}},\;
	\Gamma^0_{13}=\frac{\gamma_{c}}{2}+\gamma_{13}-i \delta_\text{p}
\end{equation}
and
\begin{equation}
	\Delta k=k_\text{p}-k_\text{c}. \label{eq:gamma-def}
\end{equation}
Using this expression for  $\rho_{31}$, we apply the slowly-varying envelope approximation for the probe field entering the vapor cell of 
length $L$ at  $\text{z}_0$ and exiting at $\text{z}_0+L$.

Denoting the initial intensity of the probe field as~$\Omega_{\text{p0}}$,
we derive an expression for  the intensity of the probe at the exit of the cell to be
\begin{align}
	\Omega_\text{p} (z_0+L)
	=&\text{e}^{-\alpha L}\Bigg(\Omega_{\text{p0}}
		-\text{i}\frac{ \alpha \Omega_\text{c}\Omega_{\mu\text{w}}
		\text{e}^{\text{i} \Delta k z_0}}{\Gamma^0_{12}}
			\nonumber\\&
		\times\frac{\text{e}^{(\alpha +i \Delta k) L}-1}{\alpha +\text{i}\Delta k}\Bigg)
\label{eq:trans}
\end{align}
with
\begin{equation}
	\alpha:=\frac{\eta \Gamma^0_{12}}{\Gamma^0_{12}\Gamma^0_{13}+|\Omega_\text{c}|^2}.
\end{equation}
By considering experimentally realistic parameters with~$L=0.05$ m and by incorporating  $^{85}$Rb hyperfine ground-state 
separation of~$3.035$~GHz,
we obtain $\Delta k=63.624$~m$^{-1}$ 
and $\alpha L \ll 1$.

At resonance with~$\delta_\text{p}=\delta_{\mu\text{w}}$
and $\delta_\text{c} = 0$,
we obtain the threshold $\gamma_\text{c}$ ($\gamma_{\text{th}}$)
by constraining the input and output intensities at 
the beginning and at the end of the vapor cell to be equal:
$\Omega_\text{p} (z_0+L) = \Omega_{\text{p0}}$.
with~$\Delta kL\sim \pi$,
this gives us
\begin{equation}
	\gamma_{\text{th}} =\frac{\Omega_{\mu\text{w}}  \Omega_\text{c}}{\Omega_{\text{p0}} \pi}.\label{eq:threshold}
\end{equation}
The analytically estimated value of~$\gamma_{\text{th}}$ at $T=0$~K, for our intensities of probe, coupling and microwave field
is around 1.62~MHz, which is in quite good agreement with the full numerically simulated value seen in our contour plot of  Fig.~\ref{fig:probeintensitychange} at $T=0$~K.

From Eq.~(\ref{eq:threshold}),
we see that $\gamma_{\text{th}}$ can be modified
by altering the intensities of coupling, probe and
microwave fields as long as the population remains predominantly in the ground state.
With increasing temperature~$T$, the thermal redistribution of 
ground-state population undermines the assumptions $\rho_{11} = 1$ and $\rho_{22} = 0$,
thereby making the dependence of 
$\gamma_{\text{th}}$ on intensities complicated.
In such regimes,
the threshold has to be obtained from a numerical plot as given in Fig.~\ref{fig:probeintensitychange}.

\section{Conclusions}
\label{sec:conclusions}
We study an atomic~$\Delta$ system interacting with optical probe and coupling fields and a microwave drive field.
Our analysis incorporates
effects of ground-state coherence decay rate and effects of thermal bath
associated with finite temperature systems
with a view to understanding regimes of probe amplification.

Our numerical results predict the existence of a threshold value for
rate of ground-state coherence decay at every temperature, below which the probe field
experiences amplification and above which it experiences absorption. We find that experimental observation of
enhancement and not amplification in such atomic~$\Delta$ systems is mainly due to the presence of ground-state decohering factors. 

We predict that enhancement is actually a precursor to probe amplification, and that amplification can be obtained if suitable reduction in ground-state decoherence 
can be achieved. Our theory thus indicates that it is possible to obtain probe amplification even for warm
$\Delta$ systems. We believe that this is an important step in experimentally obtaining phase-sensitive 
room-temperature amplification effects in equivalent~$\Delta$ system architectures.
\begin{acknowledgments}
BCS appreciates financial support from Alberta Innovates - Technology Futures, 
Natural Sciences and Engineering Research Council of Canada and 
China's 1000 Talent Program (http://1000plan.safea.gov.cn/)
and appreciates the hospitality of the Raman Research Institute during which part of this research took place.
AN thanks Professor ~G.S.~Agarwal for useful discussions.
\end{acknowledgments}
\bibliography{ref}
\end{document}